\documentclass[prd,twocolumn,showpacs,preprintnumbers,amsmath,amssymb,superscriptaddress,floatfix,nofootinbib]{revtex4}

\usepackage{graphicx}
\usepackage{amsmath}
\usepackage{amsfonts}
\usepackage{amssymb}
\usepackage{color}
\usepackage{multirow}
\usepackage[colorlinks, citecolor=blue,anchorcolor=red,menucolor=red,linkcolor=red,filecolor=red,runcolor=red,urlcolor=blue,frenchlinks=red]{hyperref}

\usepackage{subfigure}

\begin{document}

\title{Role of scalar $a_0(980)$ in the single Cabibbo Suppressed process $D^+ \to \pi^{+} \pi^{0} \eta$}

\author{Man-Yu Duan}
\affiliation{School of Physics and Microelectronics, Zhengzhou University, Zhengzhou, Henan 450001, China}

\author{Jun-Ya Wang}
\affiliation{School of Physics and Microelectronics, Zhengzhou University, Zhengzhou, Henan 450001, China}

\author{Guan-Ying Wang}
\affiliation{School of Physics and Microelectronics, Zhengzhou University, Zhengzhou, Henan 450001, China}
\affiliation{School of Physics and Electronics, Henan University, Kaifeng 475004, China}
\affiliation{International Joint Research Laboratory of New Energy Materials and Devices of Henan Province, Henan University, Kaifeng 475004, China}

\author{En Wang}
\email{wangen@zzu.edu.cn}
\affiliation{School of Physics and Microelectronics, Zhengzhou University, Zhengzhou, Henan 450001, China}

\author{De-Min Li}\email{lidm@zzu.edu.cn}
\affiliation{School of Physics and Microelectronics, Zhengzhou University, Zhengzhou, Henan 450001, China}

\begin{abstract}
Taking into account that the scalar  $a_0(980)$ can be dynamically generated from the pseudoscalar-pseudoscalar interaction within the chiral unitary approach, we have studied the single Cabibbo-suppressed 
process $D^+\to \pi^+\pi^0\eta$. We find  clear peaks of $a_0(980)^+$ and $a_0(980)^0$ in the $\pi^+\eta$ and $\pi^0\eta$ invariant mass distributions, respectively.  The predicted Dalitz plots of $D^+\to \pi^+\pi^0\eta$ also manifest the significant signals for $a_0(980)^+$ and $a_0(980)^0$ states. The uncertainties of the results due to the free parameters are also discussed. Our study shows that the process $D^+\to \pi^+\pi^0\eta$ can be used to explore the nature of the scalar $a_0(980)$, thus we encourage the experimental physicists to measure this reaction with more precision.
\end{abstract}

%\pacs{Valid PACS appear here}
% PACS, the Physics and Astronomy Classification Scheme.
% Valid PACS numbers may be entered using the \verb+\pacs{#1} command.

%\keywords{Baryons, Mesons, Resonances, Molecular states, Chiral unitary approach, Nonperturbative technique.}

\maketitle

\section{INTRODUCTION}
\label{sec:INTRODUCTION}

The studies of the charmed hadron decays are crucial to explore the strong and weak interaction effects, and to search for the $CP$ violation~\cite{Cheng:2015iom,Ebert:1983ih,Cheng:2018hwl,Cheng:1991sn,
Lu:2016ogy,Geng:2018upx,Oset:2016lyh}. Recently, the BESIII Collaboration has measured the absolute branching fraction of the $D^+\to\pi^+\pi^0\eta$ decay of $(2.23\pm0.15\pm 0.10) \times 10^{-3}$~\cite{Ablikim:2019ibo}, with much more precision than the prior measurement $(1.38\pm 0.31\pm0.16)\times10^{-3}$ of the CLEO Collaboration~\cite{Artuso:2008aa}, and no evidence of $CP$ violation is found. Although there is no significant $\rho^{+}$ and scalar $a_0(980)^{0,+}$ in the Dalitz plot of the $D^+\to\pi^+\pi^0\eta$ decay, the BESIII Collaboration has pointed out that the phase-space Monte Carlo distributions do not agree well with the data distributions due to some possible resonances, and mentioned that the amplitude analyses of this decay in the near future with large data sample at BESIII and Belle II will offer the opportunity to explore the decays of $D^+\to a_0(980)\pi$~\cite{Ablikim:2019ibo}. 

It should be stressed that the signal of the $a_0(980)$ was found in many processes. For instance,
Ref.~\cite{Xie:2014tma} has studied the decay  $D^0\to K^0_s a_0(980)$, and found a clear signal of the $a_0(980)$ in the $\pi^0\eta$ invariant mass distribution. In addition, there are  significant peaks of the $a_0(980)^0$ and $a_0(980)^+$ in the $\pi^0\eta$ and $\pi^+\eta$ invariant mass distributions of the process $D_s^+ \to \pi^{+} \pi^{0} \eta$,  as discussed in Ref.~\cite{Molina:2019udw}. Because the process $D^+\to a_0(980)\pi$ can proceed in $S$-wave, and the scalar $a_0(980)$ has a large coupling to the $\pi\eta$ channel,  we expect that there should be a sizeable signal of the $a_0(980)$ resonance if the large data sample are taken in near future.  Another example is that the analysis of the reaction $\Lambda_b\to J/\psi p \pi$ shows the existence of the hidden-charm pentaquark~\cite{Wang:2015pcn},  which is confirmed by the full amplitude analysis of the LHCb Collaboration~\cite{Aaij:2016ymb}. 

On the other hand, the nature of the low-lying light scalar resonances are still problematic, and crucial for us to understand the spectrum of the scalar mesons~\cite{Wang:2017pxm}, and there are many explanations about their nature, such as tetraquark, molecular states, and so on [see the review `Scalar mesons below 2~GeV' of Particle Data Group (PDG)~\cite{PDG2018}]. The chiral unitary approach, which provides the amplitudes of the pseudoscalar-pseudoscalar interactions, has been tested successfully in many reactions where the scalar mesons are generated dynamically. For the chiral unitary approach with the coupled channels, the potentials of the Bethe-Salpeter (BS) equation are taken from the chiral Lagrangians~\cite{Gasser:1983yg,Bernard:1995dp}, and the scattering amplitudes are obtained by solving the Bethe-Salpeter equation in all possible coupled channels that couple within $SU(3)$ to certain given quantum numbers.  Within the chiral unitary approach, the productions of the scalar $f_0(500)$, $f_0(980)$, and $a_0(980)$ have been studied in the decays of the $D^0$~\cite{Xie:2014tma}, $D^+_s$~\cite{Molina:2019udw}, $\bar{B}$ and
$\bar{B}_s$~\cite{Liang:2014ama,Liang:2014tia,Liang:2015qva,Xie:2018rqv},
$\chi_{c1}$~\cite{Liang:2016hmr},
 $\tau^-$~\cite{Dai:2018rra}, $J/\psi$~\cite{Liang:2019jtr}, $\eta_c$~\cite{Debastiani:2016ayp}, and $\Lambda_c$~\cite{Wang:2020pem}.

Up to our knowledge, there is no theoretical analyses on the process $D^+\to \pi^+\pi^0\eta$, so it is interesting to investigate the role of the $a_0(980)$ in this process.
In this work, we will perform the study of the single Cabibbo suppressed process $D^+ \to \pi^+\pi^0\eta $ taking into account the final state interactions of the meson-meson
interaction in coupled channels with the chiral unitary approach.

This paper is organized as follows. In Sec.~\ref{sec:FORMALIISM}, we will present the mechanism for the reaction of $D^+ \to \pi^{+} \pi^{0} \eta$, and in Sec.~\ref{sec:RESULTS}, we will show our results and discussions, followed by a short summary in the last section.

\section{FORMALISM}
\label{sec:FORMALIISM}

  \begin{figure}[h]
  \centering
  \subfigure[]{\includegraphics[scale=0.48]{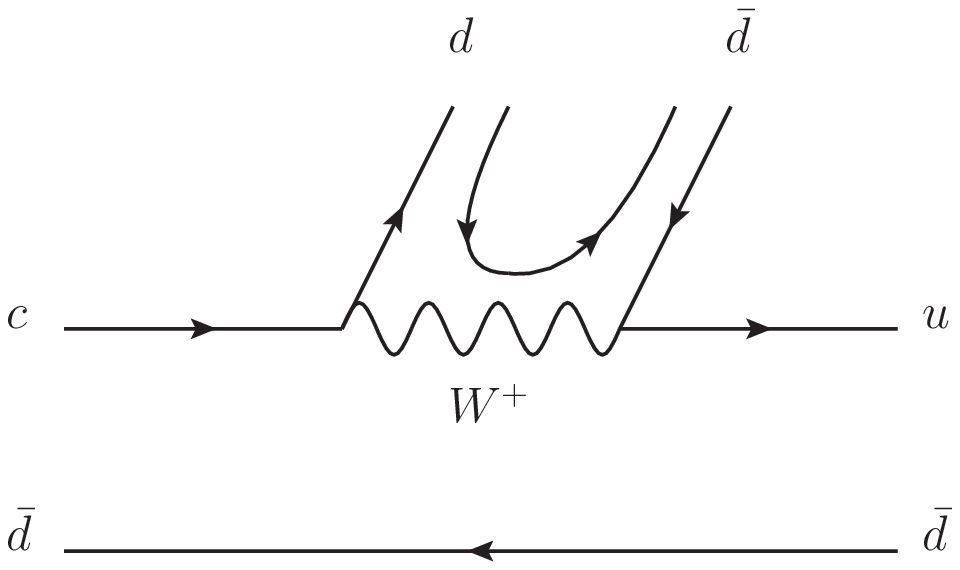}\label{fig:2a}}
  \subfigure[]{\includegraphics[scale=0.48]{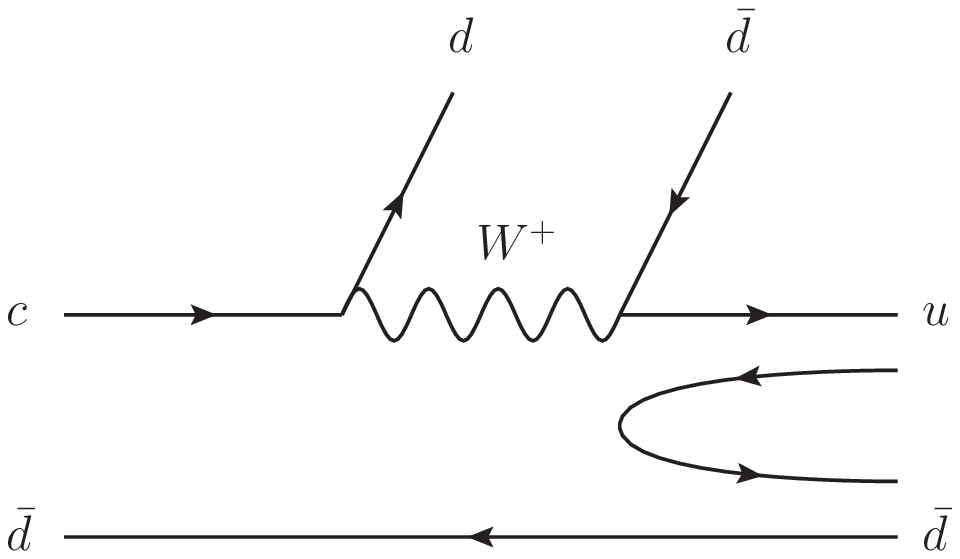}\label{fig:2b}}
  \subfigure[]{\includegraphics[scale=0.48]{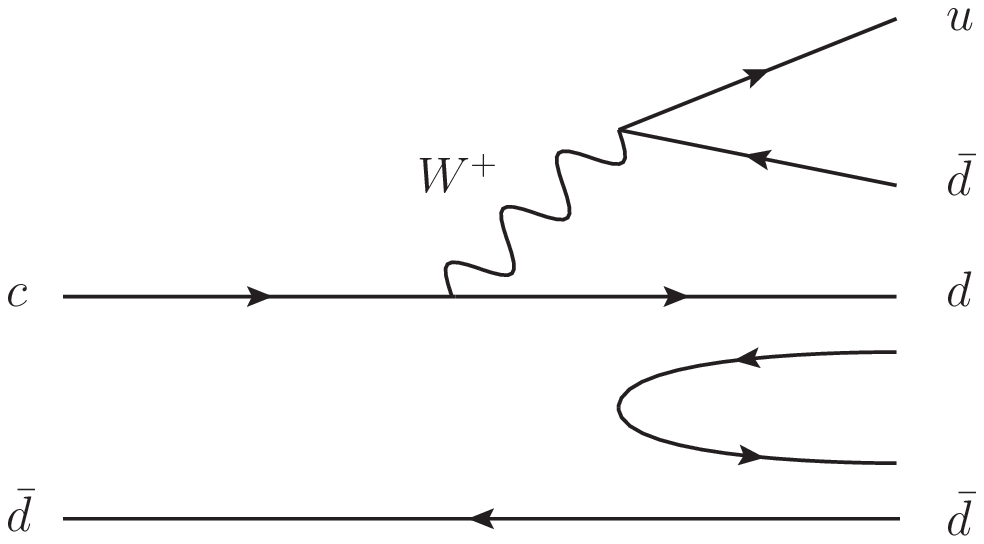}\label{fig:2c}}
  \subfigure[]{\includegraphics[scale=0.48]{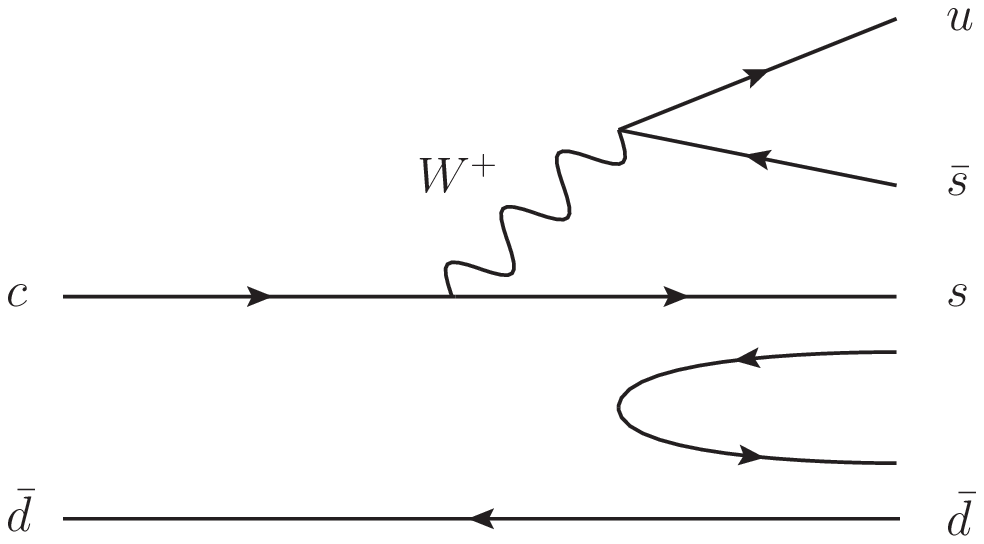}\label{fig:2d}}
  \caption{\small{Diagrammatic representation of the $D^+$ decay. (a) The internal emission of $D^+\to\pi^+d\bar{d}$ and hadronization of the $d\bar{d}$ through $\bar{q}q$ with vacuum quantum numbers. (b) The internal emission of $D^+\to\pi^0u\bar{d}$ and hadronization of the $u\bar{d}$ through $\bar{q}q$ with vacuum quantum numbers. (c) The external emission of $D^+\to\pi^+d\bar{d}$ and hadronization of the $d\bar{d}$ through $\bar{q}q$ with vacuum quantum numbers. (d)The external emission of $D^+\to K^+s\bar{d}$ and hadronization of the $s\bar{d}$ through $\bar{q}q$ with vacuum quantum numbers. }}
  \label{fig:2}
  \end{figure}

In analogy to Refs.~\cite{Xie:2014tma,Molina:2019udw}, the mechanism of the Cabibbo suppressed process $D^+ \to \pi^{+} \pi^{0} \eta$ includes three steps, weak decay, hadronization, and the final state interactions. The weak decay of the $D^+$ can happen by means of $W^+$ internal emission,  where the $c$ quark decays into a $W^+$ boson and a $d$ quark, and then $W^+$ goes to $\bar{d}$ and $u$ quarks. In order to produce the final hadrons, the $d\bar{d}$ or $u\bar{d}$ pair need to hadronize to a pair of pseudoscalar mesons with the $q\bar{q}$ ($=\bar{u}u+\bar{d}d+\bar{s}s$) produced from the vacuum as depicted in Fig.~\ref{fig:2a} or Fig.~\ref{fig:2b}, and we have,
\begin{eqnarray}
\sum_{i} d(\bar{q}_i q_{i})\bar{d}
=\sum_{i} M_{2i} M_{i2}=(M^{2})_{22} ,\label{eq:quarkdiagram1} \\
\sum_{i} u(\bar{q}_i q_{i})\bar{d}
=\sum_{i} M_{1i} M_{i2}=(M^{2})_{12} ,\label{eq:quarkdiagram2}
\end{eqnarray}
for Figs.~\ref{fig:2a} and \ref{fig:2b}, respectively,
where $M$ is the matrix in terms of the pseudoscalar mesons.
%\begin{widetext}
\begin{center}
\begin{align}
M=
\left(\begin{array}{ccc}
              \frac{\pi^0}{\sqrt{2}} +\frac{\eta}{\sqrt{3}}+\frac{\eta^\prime}{\sqrt{6}}& \pi^+ & K^+\\
              \pi^-& -\frac{\pi^0}{\sqrt{2}} +\frac{\eta}{\sqrt{3}}+\frac{\eta^\prime}{\sqrt{6}} & K^0\\
              K^-& \bar{K}^0 & - \frac{\eta}{\sqrt{3}}+\frac{2\eta^\prime}{\sqrt{6}}
      \end{array}
\right)\,.
\label{Mmatrix}
\end{align}
\end{center}
%\end{widetext}
Since the $\eta'$ has a large mass and does not play a role in the generation of the $a_{0}(980)$~\cite{Oller:1997ti}, we ignore the $\eta^\prime$ component in this work. and Eqs.~(\ref{eq:quarkdiagram1}) and (\ref{eq:quarkdiagram2}) can be re-written as,
\begin{align}\label{M22}
(M^{2})_{22}&=\pi^+\pi^-+\frac{1}{2}\pi^0\pi^0-\sqrt{\frac{2}{3}}\pi^0\eta+\frac{1}{3}\eta\eta+K^0\bar{K}^0,\\
\label{M12}(M^{2})_{12}&=\frac{2}{\sqrt{3}}\pi^+\eta+K^+\bar{K}^0.
\end{align}
Becasue the channels $\pi^+\pi^-$, $\pi^0\pi^0$, and $\eta\eta$ of Eq.~\eqref{M22} do not couple to the system of isospin $I=1$, they have no contribution in the $a_0(980)$ production, thus we have the final states after the hadronization as follows,
\begin{eqnarray}
\label{Ha}
H^{(a)}&=&V^{(a)}V_{cd}V_{ud}\left(-\sqrt{\frac{2}{3}}\pi^0\eta+K^0\bar{K}^0\right)\pi^+,\\
\label{Hb}
H^{(b)}&=&V^{(b)}V_{cd}V_{ud}\left( \frac{2}{\sqrt{3}}\pi^+\eta+K^+\bar{K}^0 \right) \left( -\frac{1}{\sqrt{2}}\pi^0\right)\nonumber \\
&=&V^{(b)}V_{cd}V_{ud}\left(-\sqrt{\frac{2}{3}}\pi^+\eta -\frac{1}{\sqrt{2}}K^+\bar{K}^0\right)\pi^0,
\end{eqnarray}
where the factor $-\frac{1}{\sqrt{2}}$ of $\pi^0$ in Eq.~(\ref{Hb}) is due to the flavor component of the $\pi^0=\frac{1}{\sqrt{2}}\left(u\bar{u}-d\bar{d}\right)$. The elements of the CKM matrix are $V_{cd}=V_{us}=-0.22534$ and $V_{ud}=V_{cs}=0.97427$~\cite{PDG2018}. The $V^{(a)}$ and $V^{(b)}$ are the factors of the production vertices of Figs.~\ref{fig:2a} and Fig.~\ref{fig:2b} containing all the dynamics, and these two factors should be similar because the weak processes of Figs.~\ref{fig:2a} and \ref{fig:2b} are the same before  hadronizations.

In addition, the mechanisms of the $W^+$ external emission shown in Figs.~\ref{fig:2c} and \ref{fig:2d} also contribute to the process $D^+\to \pi^+\pi^0\eta$. The hadronization step of Fig.~\ref{fig:2c} is the same as the one of Fig.~\ref{fig:2a}, but with an extra color factor $C$ accounting for the relative weight of the external emission mechanism with respect to the internal emission mechanism. Because the $u\bar{d}$ or $u\bar{s}$ pair from the $W^+$ decay in the external emission [Figs.~\ref{fig:2c} and \ref{fig:2d}] is constrained to form the color singlet $\pi^+$ and $K^+$ within three choices of colors, while the $u$ and $\bar{d}$ quarks from the $W^+$ decay in the internal emission [Figs.~\ref{fig:2a} and \ref{fig:2b}] have the fixed colors, the factor $C$ should be around 3~\cite{Dai:2018tgo,Zhang:2020rqr,Dai:2018nmw}. Thus, we have the possible final states for Fig.~\ref{fig:2c},
\begin{align}\label{HaaHcc}
H^{(c)}=C\times V^{(a)}V_{cd}V_{ud}\left(-\sqrt{\frac{2}{3}}\pi^+\pi^0\eta+K^0\bar{K}^0\pi^+ \right).
\end{align}
The weak process of Fig.~\ref{fig:2d} is the same as the one of Fig.~\ref{fig:2c} except for the elements of CKM matrix, thus we have,
\begin{eqnarray}
\label{HaaHcc}
H^{(d)}&=&C\times V^{(a)}V_{cs}V_{us}\left(M^2\right)_{32}K^+ \nonumber \\
&=&C\times V^{(a)}V_{cs}V_{us} \left(K^+K^-\pi^+-\frac{1}{\sqrt{2}}K^+\bar{K}^0\pi^0\right). \nonumber \\
\end{eqnarray}

Now, we sum the contributions from Figs.~\ref{fig:2a}, \ref{fig:2b}, \ref{fig:2c}, and \ref{fig:2d},
\begin{eqnarray}
H&=&H^{(a)}+H^{(b)}+H^{(c)}+H^{(d)}\nonumber\\
&=&V^{(a)} \left[  -\sqrt{\frac{2}{3}} ( 1+C+R )\pi^+\pi^0\eta + (1+C) K^0\bar{K}^0\pi^+ \right. \nonumber \\
&& \left. + C\times K^+K^-\pi^+ -\frac{1}{\sqrt{2}}(C+R) K^+\bar{K}^0 \pi^0 \right], \label{eq:fullH}
\end{eqnarray}
where the elements of the CKM matrix have been absorbed in the normalization factor $V^{(a)}$, and the $R=V^{(b)}/V^{(a)}$ stands for the relative weight of Fig.~\ref{fig:2b} with respect to Fig.~\ref{fig:2a}. Since the mechanisms of weak process of Fig.~\ref{fig:2a} and Fig.~\ref{fig:2b} are the same, we expect the $R=V^{(b)}/V^{(a)}$ to be around 1, and will calculate the results with different values of $R$. 

  \begin{figure}[tbhp]
  \centering
  \subfigure[]{\includegraphics[scale=0.47]{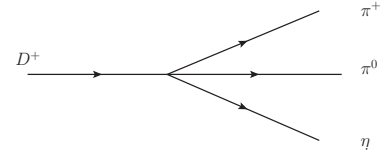}}
  \subfigure[]{\includegraphics[scale=0.47]{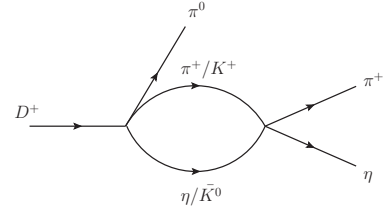}}
  \subfigure[]{\includegraphics[scale=0.47]{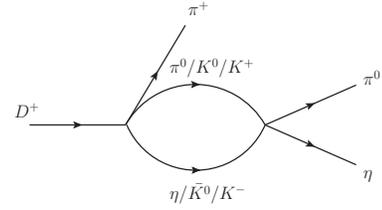}}
  \caption{\small{The mechanisms of the $D^+\to\pi^+\pi^0\eta$. (a) tree diagram, (b) the final state interaction of $\pi^+\eta$, $K^+\bar{K}^0$ and (c) the final state interaction of $\pi^0\eta$, $K^0\bar{K}^0$, and $K^+K^-$.}}
  \label{fig:3}
  \end{figure}

The full amplitude for the decay $D^+\to \pi^+\pi^0\eta$ can be easily obtained as follows,
\begin{eqnarray}
\mathcal{M}&=&V^{(a)}\left[ h_{\pi^0\pi^+\eta} + h_{\pi^0\pi^+\eta} G_{\pi^0\eta}(M_{\pi^0\eta})t_{\pi^0\eta\to \pi^0\eta }(M_{\pi^0\eta}) \right.\nonumber \\
&& +h_{K^0\bar{K}^0\pi^+}  G_{K^0\bar{K}^0}(M_{\pi^0\eta})t_{K^0\bar{K}^0\to \pi^0\eta }(M_{\pi^0\eta})\nonumber \\
&&\left. +h_{ K^+K^-\pi^+}  G_{K^+K^-}(M_{\pi^0\eta})t_{K^+K^-\to \pi^0\eta }(M_{\pi^0\eta})\right] \nonumber \\
&& + V^{(a)}\left[ h_{\pi^0\pi^+\eta} G_{\pi^+\eta}(M_{\pi^+\eta})t_{\pi^+\eta\to \pi^+\eta }(M_{\pi^+\eta}) \right. \nonumber \\
&&+ \left.  h_{ K^+\bar{K}^0\pi^0} G_{K^+\bar{K}^0}(M_{\pi^+\eta})t_{ K^+\bar{K}^0\to \pi^+\eta }(M_{\pi^+\eta}) \right], \label{eq:fullamp}
\end{eqnarray}
with $h_{\pi^0\pi^+\eta}= -\sqrt{\frac{2}{3}} ( 1+C+R )$, $h_{K^0\bar{K}^0\pi^+}=1+C$, $h_{ K^+K^-\pi^+}=C $, and $h_{ K^+\bar{K}^0\pi^0}=-\frac{1}{\sqrt{2}}(C+R)$, as taken from Eq.~(\ref{eq:fullH}).  $G_i$ is the loop function, and the transition amplitudes $t_{i\to j}$ are obtained in the Chiral unitary approach by solving the Bethe-Salepter equation,
\begin{equation}\label{BS}
T=[1-VG]^{-1}V,
\end{equation}
where $V$ is a $2\times2$ matrix with the transition potential between the isospin channels $K\bar{K}$ and $\pi\eta$. With the isospin multiplets $K=(K^+,K^0)$,
$\bar{K}=(\bar{K}^0,-K^-)$, and $\pi=(-\pi^+,\pi^0,\pi^-)$, the $2\times2$ matrix $V$ can be easily obtained as follows~\cite{Xie:2014tma},
\begin{align}\label{V}
&V_{K \bar{K} \to K \bar{K}}=-\frac{1}{4f^2}s,\\
&V_{K \bar{K} \to \pi\eta}=\frac{\sqrt{6}}{12f^2}\left(3s-\frac{8}{3}m_K^2-\frac{1}{3}m_{\pi}^2-m_{\eta}^2\right),\\
&V_{\pi\eta \to \pi\eta}=-\frac{1}{3f^2}m_{\pi}^2,
\end{align}
and the transition amplitudes $t_{i\to j}$ between charged coupled channels   can be related to the ones  between coupled channels with isospin base, 
\begin{eqnarray}
t_{K^+K^-\to\pi^0\eta}&=&-\frac{1}{\sqrt{2}}t_{K \bar{K}\to\pi\eta}^{I=1},\label{trela1}\\
t_{K^0\bar{K}^0\to\pi^0\eta}&=&\frac{1}{\sqrt{2}}t_{K \bar{K}\to\pi\eta}^{I=1},\label{trela2}\\
t_{K^+\bar{K}^0\to\pi^+\eta}&=&-t_{K\bar{K}\to\pi\eta}^{I=1}.\label{trela3}
\end{eqnarray}. 

The loop function $G_i$ of Eqs.~(\ref{eq:fullamp}) and (\ref{BS}) is given by
\begin{equation}\label{loop}
G_{i}= i \, \int \frac{d^4 q}{(2 \pi)^4} \,
\frac{1}{(q-P)^2 - m_2^2 + i \epsilon} \,
 \frac{1}{q^2 - m^2_1 + i
\epsilon},
\end{equation}
where $m_{1}$ and $m_{2}$ are the masses of the two mesons involved in the loop of the $i$th channel, and $P$ is the four-momentum of the two mesons.
The function $G_{i}$ is logarithmically divergent, there are two methods to solve this singular integral, either using the three-momentum cut-off method~\cite{Xie:2014tma,Molina:2019udw,Oller:1997ti}, or the dimensional regularization method~\cite{Guo:2016zep,AlvarezRuso:2010je,Gamermann:2006nm,Oller:1998zr,Ahmed:2020kmp,Oller:2000ma}. The choice of a particular regularization scheme does not, of course, affect our argumentation~\cite{Oller:2000ma,Doring:2011vk}. 
In this work, we use the dimensional regularization method, and the function $G_{i}$ can be re-expressed  as,
\begin{align}\label{propdr}
G_{i}=& \frac{1}{16 \pi^2} \left\{ a(\mu) + \ln
\frac{m_1^2}{\mu^2} + \frac{m_2^2-m_1^2 + s}{2s} \ln
\frac{m_2^2}{m_1^2} \right. \nonumber\\
&+ \frac{p}{\sqrt{s}} \left[ \ln(s-(m_2^2-m_1^2)+2
p\sqrt{s}) \right. \nonumber\\
&+
\ln(s+(m_2^2-m_1^2)+2 p\sqrt{s})  \nonumber\\
&
 - \ln(-s+(m_2^2-m_1^2)+2 p\sqrt{s}) \nonumber\\
&\left.\left.- \ln(-s-(m_2^2-m_1^2)+2 p\sqrt{s}) \right]\right\} 
\end{align}
with 
\begin{equation}\label{P}
p=\frac{\sqrt{(s-(m_1+m_2)^2)(s-(m_1-m_2)^2)}}{2\sqrt{s}},
\end{equation}
where $\mu$ is the scale of dimensional regularization. Following Eq.(17) in Ref.~\cite{Oller:2000fj}, we take $\mu=600$~MeV, $a(\mu)_{\pi\eta}=-1.71$, and $a(\mu)_{K\bar{K}}=-1.66$.

With the relationship of Eqs.~(\ref{trela1}-\ref{trela3}), the full amplitude of Eq.~(\ref{eq:fullamp}) can be re-written as,
\begin{eqnarray}
\mathcal{M}&=&V^{(a)}\left[ h_{\pi^0\pi^+\eta} + h_{\pi^0\pi^+\eta} G_{\pi^0\eta}(M_{\pi^0\eta})t^{I=1}_{\pi\eta\to \pi\eta }(M_{\pi^0\eta}) \right.\nonumber \\
&& +\frac{h_{K^0\bar{K}^0\pi^+}}{\sqrt{2}}   G_{K^0\bar{K}^0}(M_{\pi^0\eta})t^{I=1}_{K\bar{K}\to \pi\eta }(M_{\pi^0\eta})\nonumber \\
&& \left. -\frac{h_{ K^+K^-\pi^+}}{\sqrt{2}}  G_{K^+K^-}(M_{\pi^0\eta})t^{I=1}_{K\bar{K}\to \pi\eta }(M_{\pi^0\eta})\right]\nonumber \\
&& + V^{(a)}\left[ h_{\pi^0\pi^+\eta} G_{\pi^+\eta}(M_{\pi^+\eta})t^{I=1}_{\pi\eta\to \pi\eta }(M_{\pi^+\eta}) \right. \nonumber \\
&& - \left.  h_{ K^+\bar{K}^0\pi^0} G_{K^+\bar{K}^0}(M_{\pi^+\eta})t^{I=1}_{ K\bar{K}\to \pi\eta }(M_{\pi^+\eta}) \right]. \label{eq:fullamp2}
\end{eqnarray}

Since the amplitude of Eq.~(\ref{eq:fullamp2}) depends on two independent invariant masses $M_{\pi^0\eta}$ and $M_{\pi^+\eta}$, the double differential  width for the process $D^+\to \pi^+\pi^0\eta$ is,
\begin{eqnarray}
\frac{d^2\Gamma}{dM_{\pi^0\eta}dM_{\pi^+\eta}}&=&\frac{1}{(2\pi)^3} \frac{M_{\pi^0\eta}M_{\pi^+\eta}}{8m^3_{D^+}}{|{\cal M}|}^2.\label{eq:width}
\end{eqnarray}
We can obtain ${d\Gamma}/{dM_{\pi^0\eta}}$ and ${d\Gamma}/{dM_{\pi^+\eta}}$, by integrating Eq.~\eqref{eq:width} over each of the invariant mass variables with relations as follows,
\begin{align}\label{daliz}
(M_{\pi^0\eta}^2)_{\rm max}=(E_{\pi^0}^*+E_{\eta}^*)^2-\left(\sqrt{{E_{\pi^0}^{*2}}-m_{\pi^0}^2}-\sqrt{{E_{\eta}^{*2}}-m_{\eta}^2} \right)^2,\nonumber\\
(M_{\pi^0\eta}^2)_{\rm min}=(E_{\pi^0}^*+E_{\eta}^*)^2-\left(\sqrt{{E_{\pi^0}^{*2}}-m_{\pi^0}^2}+\sqrt{{E_{\eta}^{*2}}-m_{\eta}^2}\right)^2,
\end{align}
here $E_{\pi^0}^*$ and $E_{\eta}^*$ are the energies of $\pi^0$ and $\eta$ in the $\pi^+\eta$ rest frame, respectively,
\begin{align}\label{E}
&E_{\pi^0}^*=\frac{m_{D^+}^2-M_{\pi^+\eta}^2-m_{\pi^0}^2}{2M_{\pi^+\eta}},\nonumber\\
&E_{\eta}^*=\frac{M_{\pi^+\eta}^2-m_{\pi^+}^2+m_{\eta}^2}{2M_{\pi^+\eta}}.
\end{align}
The $\pi^0\eta$ invariant mass distribution can be obtained by interchanging the $\pi^0$ and $\pi^+$ in Eqs.~\eqref{daliz} and \eqref{E}. 

As we known, the three-body decays of charm mesons often proceed as quasi-two-body decays with  intermediate states, and the $D^+$ may decay into an $\eta$ meson and an intermediate resonance $\rho^+$, following by $\rho^+ \to \pi^+\pi^0$. Because the broad $\rho^+$ resonance provides the background contributions for the $\pi^+\eta$ and $\pi^0\eta$ invariant mass distributions, we do  the calculations by taking an energy restriction of  $M_{\pi^+\pi^0}>1$~GeV in order to eliminate the possible contribution from the intermediate $\rho^+$, as done for the process $D_s\to \pi^+\pi^0\eta$ in Refs.~\cite{Molina:2019udw,Ablikim:2019pit}.

\section{RESULTS AND DISCUSSIONS}
\label{sec:RESULTS}

\begin{figure}[htpb]
  \centering
  \includegraphics[scale=0.95]{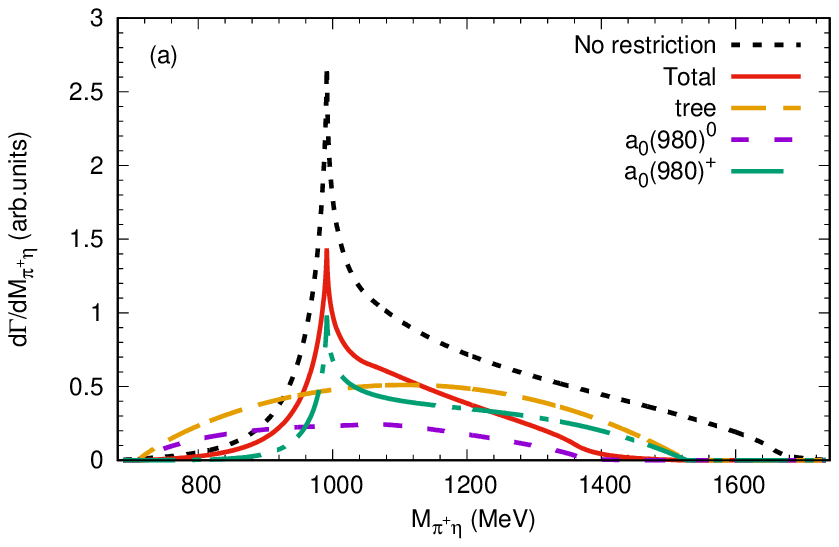}
  \includegraphics[scale=0.95]{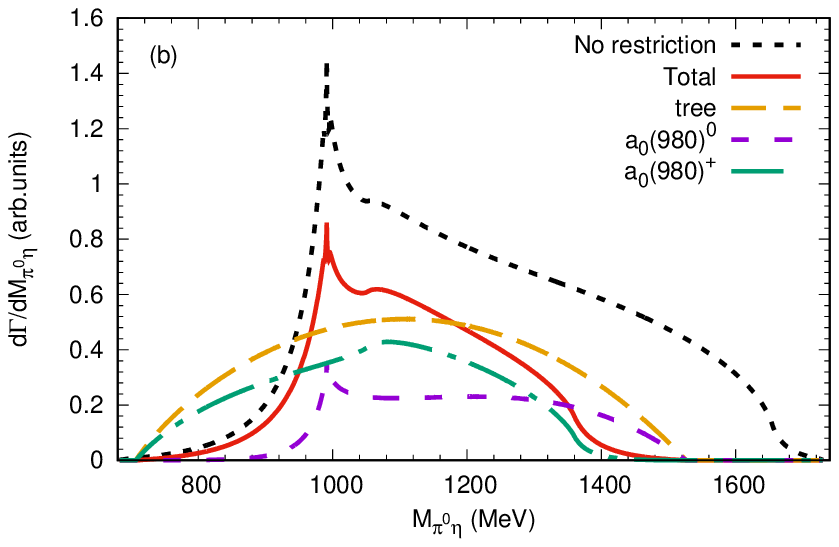}
  \caption{\small{ The $\pi^+\eta$ (a) and $\pi^0\eta$ (b) mass distributions for the process $D^+\to \pi^+\pi^0\eta$. The curves labeled as the `tree', '$a_0(980)^+$', and `$a_0(980)^0$' correspond to the contributions from the tree diagram [Fig.~\ref{fig:3}(a)], the final state interactions of $\pi^+\eta$ [Fig.~\ref{fig:3}(b)], and the final state interactions of $\pi^0\eta$ [Fig.~\ref{fig:3}(c)], respectively. The curves labeled as`Total' show the results from the total contributions of Eq.~(\ref{eq:fullamp2}) with an energy cut $M_{\pi^+\pi^0}>1$~GeV, and the `No restriction' curves stand for the results from the total contributions without the energy cut on invariant mass $M_{\pi^+\pi^0}$.}}
  \label{fig:dw}
\end{figure}

  \begin{figure}[htpb]
  \centering
  \includegraphics[scale=0.95]{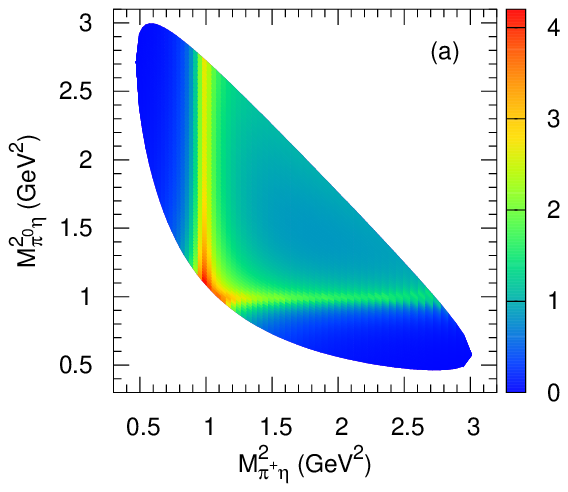}
  \includegraphics[scale=0.95]{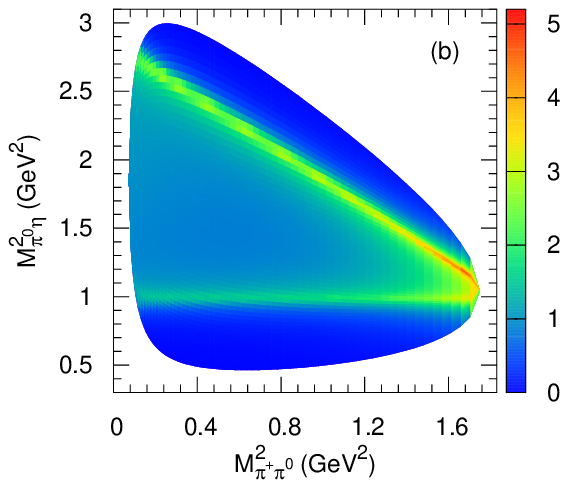}
  \caption{\small{The Dalitz plots of `$M_{\pi^0\eta}$' vs `$M_{\pi^+\pi^0}$' (a) and `$M_{\pi^0\eta}$' vs `$M_{\pi^+\eta}$' (b) for the process $D^+\to \pi^+\pi^0\eta$.}}
  \label{fig:dalitz}
  \end{figure}

  \begin{figure}[htpb]
  \centering
  \includegraphics[scale=0.8]{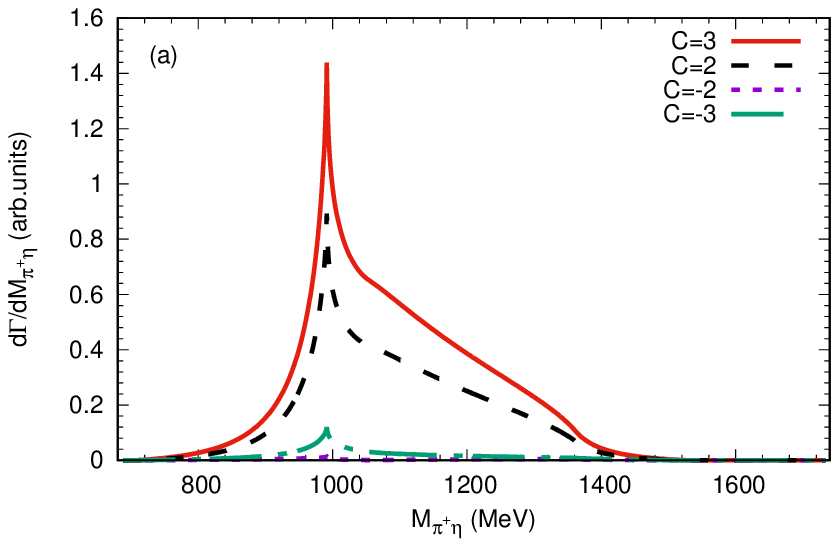}
  \includegraphics[scale=0.8]{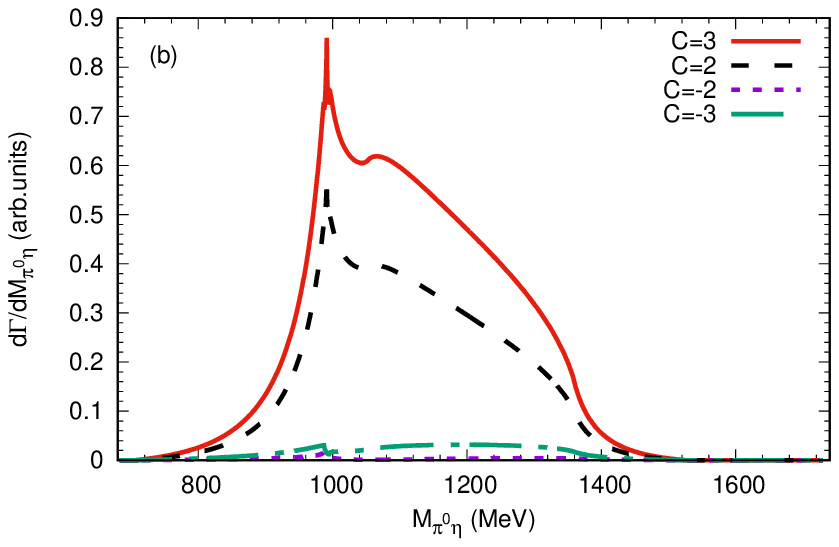}
  \caption{\small{The $\pi^+\eta$ (a) and $\pi^0\eta$ (b) mass distributions for the process $D^+\to \pi^+\pi^0\eta$ for different values of color factor $C$.}}
  \label{fig:dw_C}
  \end{figure}

  \begin{figure}[htpb]
  \centering
  \includegraphics[scale=0.8]{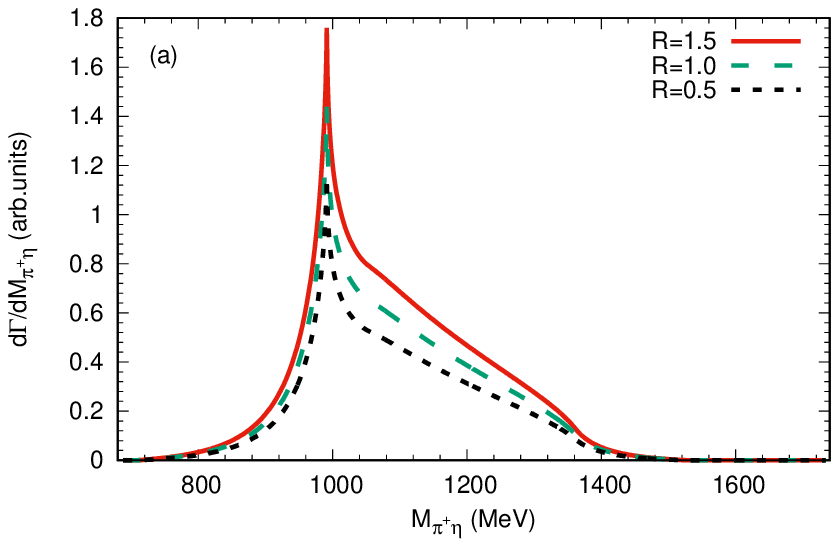}
  \includegraphics[scale=0.8]{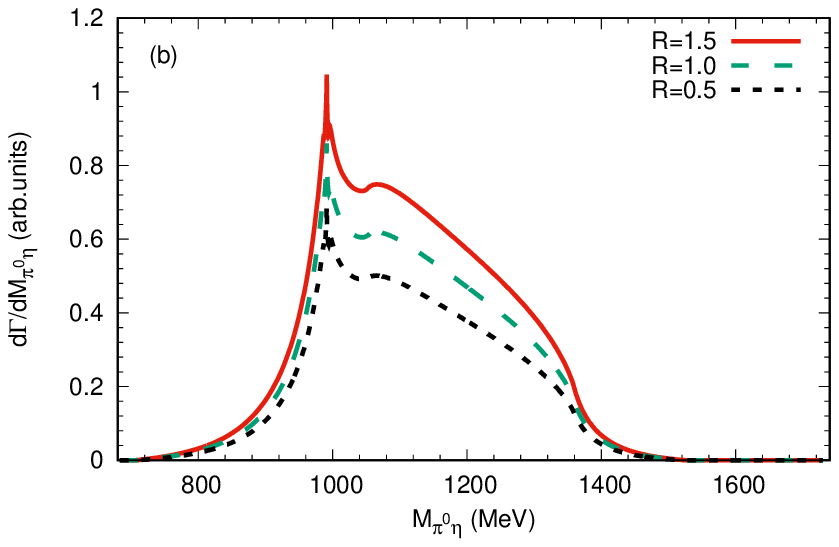}
  \caption{\small{The $\pi^+\eta$ (a) and $\pi^0\eta$ (b) mass distributions for the process $D^+\to \pi^+\pi^0\eta$ for different values of $R$.}}
  \label{fig:dw_R}
  \end{figure}

In this section, we will present our results with  above formalisms. In our model, we have three free parameters, 1) the normalization factor $V^{(a)}$ of Eq.~(\ref{eq:fullamp2}), 2) the color factor $C$, and 3) the $R=V^{(b)}/V^{(a)}$. Since the process  $D^+\to \pi^+\pi^0\eta$ has been measured by the BESIII and CLEO Collaborations~\cite{Ablikim:2019ibo,Artuso:2008aa}, which implies that the invariant mass distributions of this process are able to be measured experimentally, thus we will take $V^{(a)}=1$, and give our calculations up to an arbitrary normalization. As we discussed above, the color factor $C$ should be around 3, we will take $C=3$, and later will show the results by varying the value of $C$. Because the parameter $R=V^{(b)}/V^{(a)}$ is expected to be around 1, we take $R=1$ at the first step, and then discuss the influence from the different value of $R$. The masses of the mesons involved in this work are taken from PDG~\cite{PDG2018}.

In Fig.~\ref{fig:dw}, we show the $\pi^+\eta$ and $\pi^0\eta$ invariant mass distributions. The curves labeled as the `tree', '$a_0(980)^+$', and `$a_0(980)^0$' correspond to the contributions from the tree diagram [Fig.~\ref{fig:3}(a)], the final state interaction of $\pi^+\eta$ [Fig.~\ref{fig:3}(b)], and the final state interaction of $\pi^0\eta$ [Fig.~\ref{fig:3}(c)], respectively. The curves labeled as `Total' show the results from the total contributions of Eq.~(\ref{eq:fullamp2}) with an energy restriction of $M_{\pi^+\pi^0}>1$~GeV~\cite{Molina:2019udw,Ablikim:2019pit}. One can see a significant peak around $M_{\pi^+\eta}=980$~MeV in the $\pi^+\eta$ invariant mass distribution and a significant peak around $M_{\pi^0\eta}=980$~MeV in the $\pi^0\eta$ invariant mass distribution, which can be associated to the $a_0(980)^+$ and $a_0(980)^0$ resonances, respectively. In addition, we also present the results from the total contributions without the energy cut on invariant mass $M_{\pi^+\pi^0}$, labeled as `No restriction' curves in Fig.~\ref{fig:dw}.
By comparing the curves of `Total' with the ones of `No restriction', one can easily found that the $\rho^+$ maily contributes to the regions of $M_{\pi^+\eta}>1$~GeV and $M_{\pi^0\eta}>1$~GeV, and does not affect the peak positions of the $a_0(980)^+$ and $a_0(980)^0$.
 We also present the Dalitz plots of `$M_{\pi^0\eta}$' vs `$M_{\pi^+\pi^0}$'  and `$M_{\pi^0\eta}$' vs `$M_{\pi^+\eta}$' for the process $D^+\to \pi^+\pi^0\eta$ in Figs.~\ref{fig:dalitz}(a) and \ref{fig:dalitz}(b), which can be used to check our model in future.

As mentioned above, the color factor $C$ should be around 3, since we taken $N_c=3$ here. Indeed, the $N_c$ scaling only indicates the relative strength of the absolute values, and the relative sign is not fixed~\cite{Zhang:2020rqr}. We show the $\pi^+\eta$ and $\pi^0\eta$ mass distributions for the process $D^+\to \pi^+\pi^0\eta$ with different values of $C=3,2,-2,-3$ in Fig.~\ref{fig:dw_C}, and find that the peaks of $a_0(980)^+$ and  $a_0(980)^0$ are very clear for the positive values of $C$, and become weaker for the negative values of $C$. It should be pointed out that the positive value of $C$ is supported by the analyses of the process $\Lambda_c\to p \pi^+\pi^-$ measured by the BESIII Collaboration~\cite{Wang:2020pem} and the process $B^+\to J/\psi \omega K^+$ measured by the LHCb Collaboration~\cite{Dai:2018nmw}. 

In Fig.~\ref{fig:dw_R}, we also show the $\pi^+\eta$ and $\pi^0\eta$ mass distributions for the process $D^+\to \pi^+\pi^0\eta$  with different values of $R=1.5$, $1.0$, $0.5$. Although the strength in both $\pi^+\eta$ and $\pi^0\eta$ mass distributions become a little weaker for a smaller $R$, while the peaks of $a_0(980)^+$ and $a_0(980)^0$ are still very clear, which implies that signals of $a_0(980)^+$ and $a_0(980)^0$ do not depend on the  relative weight $R$ too much.

\section{Conclusions}
\label{sec:conc}
In this work, we have investigated the single Cabibbo suppressed process $D^+\to \pi^+\pi^0\eta$, by taking into account the pseudoscalar-pesudoscalar interaction in $S$-wave within the chiral unitary approach, where the scalar $a_0(980)$ can be dynamically generated. By including the mechanisms of the $W^+$ internal and external emissions, we have calculated the $\pi^+\eta$ and $\pi^0\eta$ mass distributions, and found the clear peaks of $a_0(980)^+$ and $a_0(980)^0$. The Dalitz plots  of `$M_{\pi^0\eta}$' vs `$M_{\pi^+\pi^0}$'  and `$M_{\pi^0\eta}$' vs `$M_{\pi^+\eta}$' are also predicted, and can be used to check our model in future. 

We have also presented the  $\pi^+\eta$ and $\pi^0\eta$ mass distributions for different values of the free parameters, the color factor $C$ and the relative weight $R$. One can find our results do not depend on the values of $R$ in the range $0.5<R<1.5$ too much.  Both the peaks of $a_0(980)^+$ and $a_0(980)^0$ are much clear for the positive values of $C$, and become weaker for the negative values of $C$.

In summary, our study indicates that  the single Cabibbo suppressed process $D^+\to \pi^+\pi^0\eta$ is suitable to explore the nature of scalar $a_0(980)$, and we encourage the experimental physicists to measure this reaction with more precision.

\begin{acknowledgments}
This work is partly supported by the National Natural Science Foundation of China under Grants No. 11505158.  It is also supported by the Key Research Projects of Henan Higher Education Institutions under No. 20A140027,  the Fundamental Research Cultivation Fund for Young Teachers of Zhengzhou University (JC202041042), and the Academic Improvement Project of Zhengzhou University.
\end{acknowledgments}

\end{document}